\documentclass[twocolumn,amsmath,amssymb,aps,pra,longbibliography,superscriptaddress,floatfix,nofootinbib]{revtex4-1}
\usepackage{physics}
\usepackage{amsmath}
\usepackage{cases}
\usepackage{url}
\usepackage{natbib}
\usepackage{textcase}
\usepackage{amssymb}
\usepackage{graphicx}
\usepackage{bm} 

\usepackage{times}
\usepackage{float}
\usepackage{multirow,microtype,color,relsize}
\usepackage{subfigure}
\usepackage[breaklinks=true]{hyperref}
\usepackage[utf8]{inputenc}
\usepackage[english]{babel}
\hypersetup{colorlinks=true,linkcolor=blue,citecolor=blue}
\hypersetup{linktocpage}
\usepackage{CJKutf8}
\usepackage{breakcites}
\usepackage{float}
\usepackage[dvipsnames]{xcolor}
\definecolor{mypine}{RGB}{1, 121, 111}

\def \be {\begin{equation}}
\def \ee {\end{equation}}

\begin{document}
\begin{CJK*}{UTF8}{gbsn}
\title{Conductivity of two-dimensional small gap semiconductors and topological insulators in strong Coulomb disorder}

\author{Yi Huang~(黄奕)}
\affiliation{School of Physics and Astronomy, University of Minnesota, Minneapolis, Minnesota 55455, USA}
\email[Corresponding author: ]{huan1756@umn.edu}

\author{Brian Skinner}
\affiliation{Department of Physics, The Ohio State University, Columbus, Ohio 43202, USA}

\author{B.\,I. Shklovskii}
\affiliation{School of Physics and Astronomy, University of Minnesota, Minneapolis, Minnesota 55455, USA}

\date{\today}

\begin{abstract}

We are honored to dedicate this article to Emmanuel Rashba on the occasion of his 95 birthday. In the ideal disorder-free situation, a two-dimensional band gap insulator has an activation energy for conductivity equal to half the band gap, $\Delta$. But transport experiments usually exhibit a much smaller activation energy at low temperature, and the relation between this activation energy and $\Delta$ is unclear. Here we consider the temperature-dependent conductivity of a two-dimensional narrow gap semiconductor on a substrate containing Coulomb impurities, mostly focusing on the case when amplitude of the random potential $\Gamma \gg \Delta$. 
We show that the conductivity generically exhibits three regimes and only the highest temperature regime exhibits an activation energy that reflects the band gap. At lower temperatures, the conduction proceeds through nearest-neighbor or variable-range hopping between electron and hole puddles created by the disorder. We show that the activation energy and characteristic temperature associated with these processes steeply collapse near a critical impurity concentration. Larger concentrations lead to an exponentially small activation energy and exponentially long localization length, which in mesoscopic samples can appear as a disorder-induced insulator-to-metal transition. 
We arrive at a similar disorder driven steep insulator-metal transition in thin films of three-dimensional topological insulators with very large dielectric constant, where due to confinement of electric field internal Coulomb impurities create larger disorder potential. Away from neutrality point this unconventional insulator-to-metal transition is augmented by conventional metal-insulator transition at small impurity concentrations, so that we arrive at disorder-driven re-entrant metal-insulator-metal transition. We also apply this theory to three-dimensional narrow gap Dirac materials.
\end{abstract}
\maketitle
\end{CJK*}

\section{Introduction}
\label{sec:intro}

\begin{figure}[t]
    \centering
    \includegraphics[width=\linewidth]{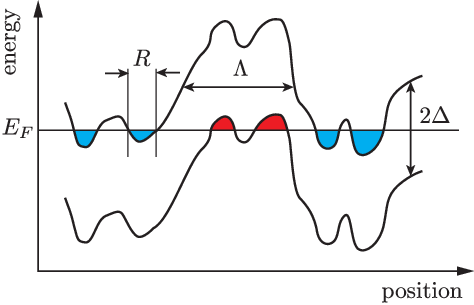}
    \caption{Schematic energy diagram of a completely compensated semiconductor with relatively weak disorder. The wavy lines show the conduction band bottom and the valence band ceiling separated by the gap $2\Delta$. Droplets of holes are shaded by red, while electron droplets are shaded by blue. Here $R$ is the size of a droplet, and $\Lambda$ is the size of a droplet cloud (puddle), which contains several droplets.}
    \label{fig:large_gap}
\end{figure}

In a band gap insulator, charged impurities often play a decisive role in determining the properties of the insulating state. Due to the long-ranged nature of the Coulomb potential that they create, such impurities produce large band bending that changes qualitatively the nature of electron conduction relative to the ideal disorder-free situation. An illustrative case is that of a three-dimensional completely-compensated semiconductor, for which positively-charged donors and negatively-charged acceptors are equally abundant and randomly distributed in space. In this case, the impurity potential has large random fluctuations, %When concentrations of donors and acceptors are equal, i. e., a semiconductor is totally compensated, randomly positioned donors and acceptors create large random potential fluctuations 
which can be screened only when the amplitude of this potential reaches $\Delta$, where $2\Delta$ is the band gap. This screening is produced by sparse electron and hole droplets, concentrated in spatially alternating electron and hole clouds (puddles)~\cite{shklovskii1972,shklovskii1984,skinner2012} (see Fig.~\ref{fig:large_gap}). 
At high enough temperatures the electrical conductivity is due to activation of electrons and holes from the Fermi level to the energy associated with classical percolation across the sample. At lower temperatures the conductivity is due to hopping between nearest neighbor puddles (NNH). At even smaller temperatures it is due to variable range hopping (VRH) between puddles. Crucially, in each of these temperature regimes the naive relation $E_a = \Delta$ is lost, where $E_a$ is the activation energy for conductivity. Only in the highest temperature regime is there a direct proportionality between $E_a$ and $\Delta$ (with a nontrivial small numeric prefactor)~\cite{skinner2012, Chen2016}; at lower temperatures the observed activation energy is non-universal and disorder-dependent~\cite{shklovskii1972,shklovskii1984}.

\begin{figure}[t]
    \centering
    \includegraphics[width=\linewidth]{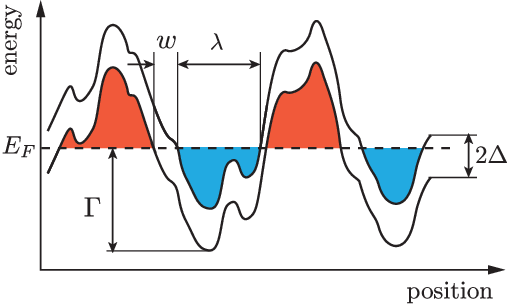}
    \caption{Schematic picture of a cross section of puddles for the case of strong disorder, $ \Gamma \gg \Delta$. The wavy lines show the conduction band bottom and the valence band ceiling separated by the gap $2\Delta$. The red shaded region above the Fermi level $E_F=0$ represents a hole puddle, while the blue shaded region below $E_F$ represents an electron puddle. $\Gamma$ is the amplitude of the disorder potential, $\lambda$ is the screening length, and $w$ is the width of the barrier between neighboring puddles. } 
    \label{fig:puddles}
\end{figure}

In this paper we consider a similar problem in two dimensions. Specifically, we consider a two-dimensional small band gap semiconductor resting on a thick substrate with a three-dimensional concentration of randomly-positioned impurities and focus on the case when $\Gamma \gg \Delta$ (see Fig.~\ref{fig:puddles}). We derive the temperature dependence of the electrical conductivity across all temperature regimes and show that the observed activation energy of the conductivity can be very small.

Understanding the relation between the energy gap and the observed activation energy for transport is of crucial importance for studying a variety of new 2D electron systems.
For example, recent studies of
%of the integer and fractional quantum Hall effects \cite{Giesbers2007, Gamez2013, Ghahari2011}, 
2D topological insulators (TIs) \cite{Olshanetsky2015, Kvon2020, Pan2020}, films of 3D TIs~\cite{nandi2018,chong2021,checkelsky2012,chang2013,he2013,mogi2015,liguozhang2017,wang2018,fox2018,moon2019,rosen2019,okazaki2020,rodenbach2021,fijalkowski2021,ferguson2021,rosen2021,okazaki2022}, bilayer graphene (BLG) with an orthogonal electric field~\cite{zou2010,taychatanapat2010} and twisted bilayer graphene (TBG)~\cite{Serlin_Intrinsic_2020, Stepanov_untying_2020, Park_Flavour_2021, cao_correlated_2018, Cao_unconventional_2018} use the transport activation energy as a way of characterizing small energy gaps. In all these cases the observed activation energy is much smaller than the energy gap that is expected theoretically or measured through local probes like optical absorption or scanning tunneling microscopy. 

Here, we show that there is indeed no simple proportionality between the energy gap and the activation energy except at the highest temperature regime, which is likely irrelevant for many experimental contexts. Instead, we find a wide regime of temperature and disorder strength for which the activation energy is dramatically smaller than the energy gap. At the lowest temperatures the conductivity follows the Efros-Shklovskii (ES) law rather than an Arrhenius law, and this dependence can give the appearance of a small activation energy.

Let us dwell on two likely applications of our theory. First, our results may be especially relevant for ongoing efforts to understand the energy gaps arising in TBG at certain commensurate fillings of the moir\'{e} superlattice \cite{Serlin_Intrinsic_2020, Stepanov_untying_2020, Park_Flavour_2021, cao_correlated_2018, Cao_unconventional_2018}. Such gaps apparently arise from electron-electron interactions, but the observed activation energies of the maximally-insulating state are typically an order of magnitude smaller than the naive interaction scale (see, e.g., Refs.~\onlinecite{Stepanov_untying_2020, Park_Flavour_2021}), and they vary significantly from one sample to another. Scanning tunneling microscopy studies also suggest a gap on the order of ten times larger than the observed activation energy \cite{Xie_spectroscopic_2019, choi2021interactiondriven}. The theory we present here offers a natural way to interpret this discrepancy.

Second, our theory can be applied to the huge body of experimental work on thin films of 3D TI, where the surface electrons have a small gap $2\Delta$ due to hybridization of the surface states of two surfaces~\cite{nandi2018,chong2021}, or due to intentionally introduced magnetic impurities~\cite{checkelsky2012,chang2013,he2013,mogi2015,liguozhang2017,wang2018,fox2018,moon2019,rosen2019,okazaki2020,rodenbach2021,fijalkowski2021,ferguson2021,rosen2021,okazaki2022}. Understanding the origin of the small apparent activation energy $E_a \ll \Delta$ is crucial for achieving metrological precision of the quantum anomalous Hall effect~\cite{yu2010,chang2013,jinsongzhang2013,mogi2015,fox2018,okazaki2020,rodenbach2021,fijalkowski2021,ferguson2021,rosen2021,okazaki2022,chang2022quantum} and the quantum spin Hall effect~\cite{liu2010,lu2010,linder2009,chong2021}.

The model we consider is a two-dimensional semiconductor with band gap $2\Delta$ atop a substrate with a three-dimensional concentration $N$ of random sign charged impurities. 
We assume that the semiconductor has a gapped Dirac dispersion law 
\begin{align} \label{eq:spectrum}
\epsilon^2(\vb{k}) = (\hbar v k)^2 + \Delta^2. 
\end{align}
We are mostly interested in the case when the amplitude $\Gamma$ of spatial fluctuations of the random potential satisfies
$\Gamma \gg \Delta$, so that electron and hole puddles occupy almost half of the space each and are separated by a small insulating gap which occupies only a small fraction of the space (see Fig.~\ref{fig:puddles}). 
This system is an insulator because in 2D neither electron nor hole puddles percolate, and they are disconnected from each other. Throughout this paper we mostly focus on the case of zero chemical potential, for which electron and hole puddles are equally abundant and the system achieves its maximally insulating state.
We argue that this situation is likely realized in the experiments of Refs.~\cite{
%Giesbers2007, Gamez2013, Ghahari2011, 
Olshanetsky2015, Kvon2020, Pan2020, zou2010, taychatanapat2010, Serlin_Intrinsic_2020, Stepanov_untying_2020, Park_Flavour_2021, cao_correlated_2018, Cao_unconventional_2018, Xie_spectroscopic_2019, choi2021interactiondriven,  nandi2018,chong2021,checkelsky2012,chang2013,he2013,mogi2015,liguozhang2017,wang2018,fox2018,moon2019,rosen2019,okazaki2020,rodenbach2021,fijalkowski2021,ferguson2021,rosen2021,okazaki2022}.

The remainder of this paper is organized as follows. In the following section we first summarize our main results for the temperature-dependent conductivity. Secs.~\ref{sec:fractal} and ~\ref{sec:hopping} concentrate on the case $\Gamma \gg \Delta$ illustrated by Fig.~\ref{fig:puddles}.
In Sec.~\ref{sec:fractal} we start from reviewing the fractal geometry of two-dimensional puddles and then calculate the action accumulated by electrons tunneling across the gap between two neighboring fractal metallic puddles, the corresponding localization length, and the critical value of the ratio $\Gamma/\Delta$, at which crossover to weak localization takes place. In Sec.~\ref{sec:hopping} we calculate the hopping conductivity for the case $\Gamma \gg \Delta$. 

In Sec.~\ref{sec:lowdisorder} we study the illustrated by Fig.~\ref{fig:large_gap} case where the impurity concentration $N$ is lower and present the parameters of NNH and VRH as functions of $N$. Section \ref{sec:PD} studies what happens when the Fermi level moves away from the neutrality point.
We arrive at the ``phase diagram'' of the re-entrant metal - insulator -  metal transition.
Section \ref{sec:TI} deals with the generalization of our results to thin TI films. 
Because of large interest to such films~\cite{liu2010,lu2010,linder2009,zhang2010a,sakamoto2010,zhang2013,kim2013,nandi2018,chong2021,chen2010,yu2010,xu2012,checkelsky2012,jinsongzhang2013,he2013,mogi2015,ye2015,liguozhang2017,wang2018,fox2018,moon2019,rosen2019,tokura2019,okazaki2020,deng2020,rodenbach2021,lu2021,fijalkowski2021,ferguson2021,rosen2021,okazaki2022,chang2022quantum}, in this section we add a fair amount of numerical estimates.
In Sec.~\ref{sec:3D} we briefly return to the problem of three-dimensional, completely-compensated semiconductors with a gapped Dirac dispersion, and extend the previous theory~\cite{shklovskii1972,shklovskii1984,skinner2012} to the case when disorder potential fluctuations exceed $\Delta$. We again arrive at a re-entrant metal - insulator - metal transition away from the neutrality point. We close in Sec.~\ref{sec:conclusion} with a summary and conclusion. Some results of this paper are published in its shorter version~\cite{huang2021conductivity}.

\section{Summary of results }
\label{sec:summary}
\begin{figure}[t]
    \centering
    \includegraphics[width=0.9\linewidth]{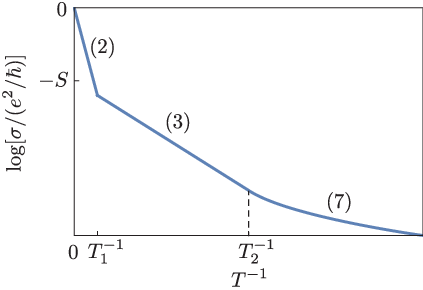}
    \caption{Logarithm of the dimensionless conductivity $\sigma/(e^2/\hbar)$ as a function of the inverse temperature $T^{-1}$ in the case $1\ll \Gamma/\Delta \ll (\Gamma/\Delta)_c$. At high temperature $T > T_1$, the conductivity has activation energy $\Delta$. At intermediate temperature $ T_2 < T < T_1$, the conductivity is dominated by NNH. At low temperatures $T < T_2$, NNH is replaced by ES VRH. Numbers adjacent to different parts of the line show corresponding equations. Temperatures $T_1$ and $T_2$ are given by Eqs.~\eqref{eq:t_1} and ~\eqref{eq:t_2}.}
    \label{fig:conductivity}
\end{figure}

Let us start from the strong disorder case $\Gamma \gg \Delta$ illustrated in Fig.~\ref{fig:puddles}. When the typical tunneling transparency $P=\exp(-S)$
of the insulating barrier separating neighboring puddles is small (the action $S$ in units of $\hbar$ is large), one can envision a sequence of three mechanisms of activated transport replacing each other with decreasing temperature, as in a lightly doped wide gap semiconductor~\cite{shklovskii1984}.
This three-mechanism sequence is illustrated in Fig.~\ref{fig:conductivity}.
At relatively large temperature $T$ electrons and holes can be activated from the Fermi level to the percolation level (i.e., the classical mobility edge). 
Thus, the conductivity at such large temperatures is given\cite{huang2021conductivity} by 
\begin{align}  \label{eq:sigmaD}
    \sigma = \sigma_1 \exp(-\Delta/T), \hspace{5mm} (T_1 \ll T \ll \Delta) 
\end{align}
with the prefactor $\sigma_1 \sim e^2/\hbar$. Here and everywhere in this paper we use energy units for the temperature $T$ (absorbing $k_B$ in its definition).

At lower temperatures this mechanism yields to the nearest-neighbor hopping (NNH) of electrons between electron and hole puddles near the Fermi level.
Similarly to the case of granular metals~\cite{chen2012, zhang2004}, the activation energy of such hopping is determined by the typical puddle charging energy $E_C$
\begin{align} \label{eq:sigma_NNH}
\sigma=\sigma_2 \exp(-E_C/T), \hspace{5mm} (T_2 \ll T \ll T_1).
\end{align}
Here the prefactor $\sigma_2\sim (e^2/\hbar) \exp(-S) \ll (e^2/\hbar)$. We show below that 
\begin{align} \label{eq:ec_2}
E_C =\alpha^2 \Delta (\Delta/\Gamma)^{4/3} = \alpha^2 \Delta (N_0/N)^{4/9}\ll \Delta. 
\end{align}
Here $\alpha = e^2/(\kappa \hbar v)$ is the analog of the fine structure constant and $\kappa$ is the dielectric constant of the substrate. With the standard semiconductor value $v\sim 10^{6}$ m s$^{-1}$; and with $\kappa=4$ for SiO$_2$, 11 for insulating GaAs, 20 for HfO$_2$ and 1000 for PbTe; $\alpha$ can vary from 1 to $10^{-3}$. Below in our theory we use $\alpha$ as a small parameter, $\alpha \ll 1$, but our results are semi-quantitatively correct  at $\alpha=1$.
In Eq.~\eqref{eq:ec_2} we also used derived below equation
\begin{align} \label{eq:gammadelta}
\Gamma/\Delta = (N/N_0)^{1/3}, 
\end{align}
where characteristic concentration 
\begin{align} 
N_0(\Delta) =\alpha^2\kappa^3\Delta^{3}e^{-6}.
\label{eq:N0}
\end{align}
In Eqs.~\eqref{eq:ec_2}, \eqref{eq:gammadelta} and \eqref{eq:N0} and everywhere below we use a scaling approach and omit all numerical coefficients.

At even lower temperatures NNH crosses over to VRH obeying the Efros-Shklovskii (ES) law 
\begin{align} \label{eq:sigma_ES}
\sigma=\sigma_3 \exp[-\left(T_{\rm ES}/T\right)^{1/2}], \hspace{5mm} (T \ll T_2),
\end{align}
with $\sigma_{3} \sim e^2/\hbar$. 
We show below that in this regime
\begin{align} \label{eq:tes}
T_{\rm ES} = \alpha \Delta (\Delta/\Gamma)^{37/9} = \alpha \Delta (N_0/N)^{37/27} \ll \Delta,
\end{align}
and the temperatures associated with the crossover between different regimes are 
\begin{align} \label{eq:t_1}
T_1 = \alpha\Delta(\Gamma/\Delta)^{34/9} = \alpha\Delta(N/N_0)^{34/27},
\end{align}
\begin{align} \label{eq:t_2}
T_2 = \alpha^3 \Delta (\Gamma/\Delta)^{13/9}=\alpha^3 \Delta(N/N_0)^{13/27}.
\end{align}

Above we dealt with large impurity concentration $N > N_0$, which corresponds to $\Gamma \gg \Delta$. In Sec.~\ref{sec:lowdisorder} of this paper we study the case $N<N_0$, for which small and sparse electron and hole droplets are able to screen the random potential of impurities, as in the three-dimensional case studied in Refs.~\cite{shklovskii1972,shklovskii1984,skinner2012} (see Fig.~\ref{fig:large_gap}). We briefly review our results here. In this case, conductivity is also due to the three-mechanism sequence. The high temperature mechanism is due to free electrons activated by energy $\Delta$ and the low temperature mechanism is ES VRH with new $T_{\rm ES}=\alpha \Delta$. 
However, activated NNH of the intermediate temperature regime is replaced by the new hopping mechanism, which we call NNH-VRH hybrid (H) mechanism. 
It  works in the new temperature interval $T_2'\ll T \ll T_1'$. 
We study this hybrid mechanism in Sec.~\ref{sec:lowdisorder}. 
Here we only want to give a hint to its physics and origin of the term ``hybrid''. 
Indeed, if we focus on puddles this is hopping between nearest neighbor puddles. However, each puddle has many droplets. Therefore, focusing on droplets we deal with VRH.

Optimization of Miller-Abrahams resistor network~\cite{shklovskii1984} of all available pairs of droplets of two adjacent puddles leads to the new H-mechanism conductivity:
\begin{align} \label{eq:sigma_hybrid}
\sigma=\sigma_2 \exp[-\left(T_H/T\right)^{5/9}],
\end{align}
where
\begin{align} \label{eq:t_H}
T_H = \alpha^{6/5}(N/N_0)^{1/5}\Delta,
\end{align}

\begin{figure}[t]
    \centering
    \includegraphics[width=0.9\linewidth]{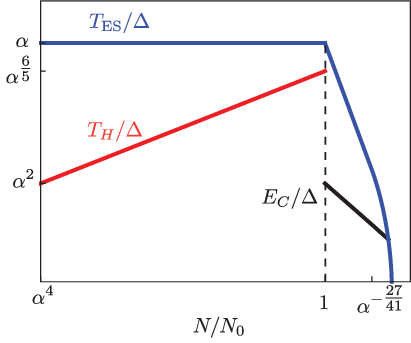}
\caption{Schematic log-log plots of characteristic energies of three kinds of hopping conductivity.
The characteristic temperature of ES law $T_{\rm ES}$ (blue line), the activation energy of NNH, $E_C$ (black solid line) and the characteristic temperature of hybrid conductivity $T_H$ (red line) are shown as functions of the dimensionless impurity concentration $N/N_0 = (\Gamma/\Delta)^3$. 
The left part of the plot where $N/N_0 < 1$ corresponds to Eqs.~\eqref{eq:tes_d} and \eqref{eq:t_H}, while the right part at $N/N_0 > 1$ corresponds to Eqs.~\eqref{eq:tes} and \eqref{eq:ec_2}. In the horizontal axis $N/N_0 = N_c/N_0 = \alpha^{-27/41}$ corresponds to $\Gamma/\Delta = (\Gamma/\Delta)_c$ given by Eq.~\eqref{eq:critical}. At this point $T_{\rm ES} = \alpha^{87/41} \Delta$. When $\Gamma/\Delta > (\Gamma/\Delta)_c$ the localization length $\xi$ increases exponentially and $T_{\rm ES}$ decreases exponentially.}
%Numbers adjacent to different parts of the line show corresponding equations. \YH{Numbers to be put later when the equation numbers condense.} }
    \label{fig:energy_n}
\end{figure}

Our results for $T_{\rm ES}$, $E_C$ and $T_H$ as a function of the dimensionless impurity concentration $N/N_0 = (\Gamma/\Delta)^3$
are summarized in Fig.~\ref{fig:energy_n}. So far in this section we did not touch the specifics of TI films discussed below, but qualitatively the results for both cases are similar.

However, such three-mechanisms sequences are not observed in most experiments~\cite{Olshanetsky2015, Kvon2020, Pan2020, zou2010, taychatanapat2010, Serlin_Intrinsic_2020, Stepanov_untying_2020, Park_Flavour_2021, cao_correlated_2018, Cao_unconventional_2018, Xie_spectroscopic_2019, choi2021interactiondriven,  nandi2018,chong2021,checkelsky2012,chang2013,he2013,mogi2015,liguozhang2017,wang2018,fox2018,moon2019,rosen2019,okazaki2020,rodenbach2021,fijalkowski2021,ferguson2021,rosen2021,okazaki2022}. 
Instead, experiments tend to report an activated conductivity with activation energy much smaller than $\Delta$.

Here we suggest a possible explanation for such low activation energies. We show below that at $\Gamma/\Delta > \alpha^{-9/41}$ electrons are not localized in single puddles and the first two regimes of conductivity are absent. The only remaining mechanism is the ES VRH with very small $T_{\rm ES}$. This means that the low temperature ``local activation energy'' is much smaller than $\Delta$. 

An alternative explanation involves the intermediate temperature regimes at $1 < N/N_0 < \alpha^{-27/41}$ or at $\alpha^4 < N/ N_0 < 1$. In this case  all three mechanisms are present in principle, but at small enough $\alpha$ the intervals $T_2 < T  < T_1$ and $T_2' < T  < T_1'$ can be large and the observed activation energy can be very small. 
However, the theoretical prefactor of the NNH conductivity and hybrid hopping conductivity $\sigma_2 \ll e^2/\hbar$. This agrees with some experiments~\cite{rosen2019,fijalkowski2021}, but contradicts to other ones, where prefactor close to $e^2/\hbar$ was observed~\cite{fox2018,rodenbach2021}. For such experiments the ES mechanism seems to provide a better explanation.

\section{Fractal geometry of puddles and tunneling action} 
\label{sec:fractal}

Let us start from the brief review geometrical fractal properties of 2D puddles at $\Gamma \gg \Delta$~\cite{isichenko1992}.
The characteristic size (diameter) of a puddle is given by 
\begin{equation}\label{eq:a}
    a = \lambda (\Gamma / \Delta)^{\nu},
\end{equation}
where $\nu = 4/3$ and $\lambda$ is the the electron screening radius. The perimeter of a puddle reads
\begin{align} \label{eq:L}
    L = a (\Gamma / \Delta) = \lambda (\Gamma / \Delta)^{\nu + 1}.
\end{align}
The perimeter $L$ is parametrically longer than the diameter $a$ because puddles have many ``fingers'', which are interlocked with other fingers of neighboring puddles [see Fig.~\ref{fig:fingers}].
The area of a puddle is given by
\begin{align}\label{eq:A}
    A = \lambda^2 (\Gamma/\Delta)^{2\nu - \beta},
\end{align}
where $\beta = 5/36$.
The separation distance between nearest-neighbor electron and hole puddles is
\begin{align}\label{eq:w}
    w = \lambda \Delta /\Gamma.
\end{align}

In order to estimate $\Gamma$ and $\lambda$ we can use the self-consistent theory of Ref.~\cite{skinner2013a}, which dealt with the disorder potential at the surface of a bulk TI created by charged impurities with three-dimensional concentration $N$. 
In our case the substrate plays the role of the TI bulk and the two-dimensional semiconductor plays the role of the TI surface.
The band gap $\Delta$ that exists in our case is not important when $\Gamma \gg \Delta$. 
To begin, we relate $\Gamma$ to $\lambda$ as the typical Coulomb energy created by charge fluctuations in a volume $\lambda^3$:
\begin{align}\label{eq:gamma0}
\Gamma = \frac{e^2}{\kappa\lambda} (N\lambda^3)^{1/2}.
\end{align}
This relation leads to a typical 2D density of states (DOS)
\begin{align}\label{eq:dos}
    g = \kappa^2 \alpha^2 \Gamma/e^4,
\end{align}
which in turn leads to the screening radius 
\begin{align}\label{eq:lambda1}
\lambda = \frac{\kappa}{e^2 g} = \frac{e^2}{\alpha^2\kappa \Gamma}.
\end{align}
Solving Eqs.~\eqref{eq:gamma0} and \eqref{eq:lambda1} for $\Gamma$ and $\lambda$ we get~\cite{skinner2013a}
\begin{align}\label{eq:gamma}
    \Gamma & = \frac{e^2 N^{1/3}}{\kappa \alpha^{2/3}}, \\
    \lambda & = \alpha^{-4/3} N^{-1/3}.\label{eq:lambda}
\end{align}

Let us now estimate the dimensionless action $S$ (the action in units of $\hbar$) electron accumulate tunneling between nearest neighbor fractal metallic electron and hole puddles separated by narrow insulating gaps (see Fig.~\ref{fig:puddles}). 
The value of $S$ is determined by the tunneling length $r=\Delta/eE$ in the spatially-varying electric field $E$ created by impurities:
\begin{align}\label{eq:action}
S = \frac{r \Delta}{\hbar v} = \frac{\Delta^2}{eE\hbar v}.
\end{align}
It is tempting to use $\Gamma/e\lambda$ for $E$ and arrive at $S = w\Delta/\hbar v = \alpha^{-1} (\Delta/\Gamma)^2$.
However, the electric field has strong fluctuations at short distances, so the typical electric field depends on the tunneling distance $r$. Since a cube of size $r$ has a typical excess charge $\sqrt{Nr^3}$, the typical electric field associated with the length scale $r$ is $E(r)= e(Nr^3)^{1/2}/\kappa r^2$, which grows with decreasing $r$. %So we should look at small enough $r$. 
Also, due to the large perimeter length $L$ of puddles we can find rare places where the random electric field is created by a larger-than-average number of excessive charges, $M\gg (Nr^3)^{1/2}$, leading to even larger electric field $E(r)= e M/\kappa r^2$. Below we find the optimal values of $M$ and $r$ which determine $S$, and we arrive at a value of $S$ value much smaller than %Eq.~\eqref{eq:action_w1}
the naive estimate $S = \alpha^{-1} (\Delta/\Gamma)^2$. 
Our optimization procedure is a mesoscopic version of the optimization used in the theory of the interband absorption of light in compensated three-dimensional semiconductors~\cite{shklovskii1984,shklovskii1970}. It is also similar to the theory of fluctuation-induced excess currents in reverse biased $p$-$n$ junctions~\cite{raikh1985}. 

Below we use $S$ to calculate the localization length $\xi$ that determines hopping transport. Thus, we are interested in fluctuations of electric field which, although rare, happen roughly once at every interface between nearest-neighboring puddles. Thus,
\begin{align}\label{eq:m}
    (L/\lambda) \exp[-\frac{M^2}{Nr^3}] = 1.
\end{align}
Here we use the Gaussian probability of finding net charge $M$
in a cube of size $r$. For tunneling across the gap $2\Delta$ we need the potential difference across the cube $Me^2/\kappa r=\Delta$. In other words, $r=r(M) = M e^2 /\kappa\Delta$. Substituting $r(M)$ into Eq.~\eqref{eq:m} and solving for $M$ gives
\begin{align}
    M = \frac{\alpha^{-2} (\Delta/\Gamma)^3}{\ln[(\Gamma/\Delta)^{7/3}]}, 
\end{align}
which at $\Gamma \gg \Delta$ corresponds to $r(M) \ll w \ll \lambda$.

Substituting the electric field $E = M e / \kappa r^{2}(M)$ into the tunneling action Eq.~\eqref{eq:action} we have
\begin{align}\label{eq:action_meso}
    S = \frac{\alpha^{-1} (\Delta/\Gamma)^3}{\ln[(\Gamma/\Delta)^{7/3}]} \simeq \alpha^{-1} (\Delta/\Gamma)^{34/9}.
\end{align}
In the last step we used the power-law approximation $\ln x=x^{1/3}$ valid for $x \in (3,100)$ with accuracy better than 30\% . 

Now we can calculate the electron localization length, $\xi$, which we need below to calculate the hopping conductivity. 
After each tunneling through the gap, electron spreads by distance $a$.
This means that at a large distance $x$ the electron accumulates an action $Sx/a = x/\xi$ where 
\begin{align}\label{eq:xi} 
    \xi = a/S = \alpha a (\Gamma/\Delta)^{34/9}.
\end{align}
The fast decrease of $S$ with growing $\Gamma/\Delta$ leads to growth of dimensionless conductance $G$ between two neighboring puddles 
\begin{align}\label{eq:g} 
 G = (L/\lambda)\exp(-S),
\end{align}
so that we get $G=1$ at some critical value $(\Gamma/\Delta)_c$. 

Substituting Eq.~\eqref{eq:action_meso} into Eq.~\eqref{eq:g} and setting $G=1$,  in terms of power law we arrive at the critical point~\footnote{In the limit of $\alpha \to 0$, the asymptotic expression to first order reads $(\Gamma/\Delta)_c = \alpha^{-1/3}[\ln(\alpha^{-1})]^{-2/3}$.}
\begin{align} \label{eq:critical} 
    (\Gamma/\Delta)_c = \alpha^{-9/41},
\end{align}
valid for $\alpha \in (1.2 \times 10^{-4}, 0.12)$.
This range of $\alpha$ is obtained by substituting Eq.~\eqref{eq:critical} into the requirement for the argument of the logarithm $(\Gamma/\Delta)^{7/3} \in (3, 100)$~\footnote{Note that this mesoscopic optimization method based on Eq.~\eqref{eq:m} is self-consistent if $\Gamma/\Delta < (\Gamma/\Delta)_c$ (or $G < 1$), so that $e^S > L/\lambda$.}.

At larger $\Gamma / \Delta$ the localization length grows exponentially as $\xi = a e^G$.
This leads to dramatic growth of the conductivity, namely to insulator - almost metal transition if the sample size is much larger than $\xi$. For a very small sample, this $(\Gamma/\Delta)_c$ effectively plays the role of the critical value of insulator-metal transition.

\section{Hopping conductivity}
\label{sec:hopping}

The character of the conductivity of our system apparently changes at $\Gamma/\Delta = (\Gamma/\Delta)_c$. 
At moderate disorder when $1 < \Gamma/\Delta < (\Gamma/\Delta)_c$ electrons are well localized within a puddle and the temperature dependence of the conductivity follows the three-mechanism sequence discussed in the Introduction. 
In strong disorder case $\Gamma/\Delta > (\Gamma/\Delta)_c$, the localization length $\xi \gg a$ and at all temperatures the conductivity is due to ES VRH with very small $T_{\rm ES}$ and the prefactor $\sigma_0 \sim e^2/\hbar$. In the limited temperature range it can look like activated transport with very small activation energy. 

Below we concentrate on the three-mechanism sequence case, when with decreasing temperature the activated conductivity with activation energy $\Delta$ is replaced first by NNH and then by ES VRH. This case reminds systems of densely packed  metallic granules separated by a thin insulator with Coulomb impurities and we can follow the calculation of their conductivity~\cite{zhang2004,chen2012}.  

\begin{figure}[t]
    \centering
    \includegraphics[width=0.8 \linewidth]{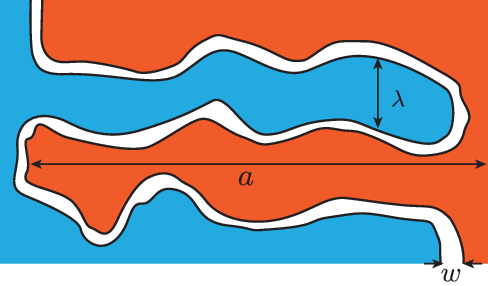}
    \caption{Schematic picture of interlocked ``fingers'' of neighboring puddles. Here the length of ``fingers'' $a$ is of order of the puddle diameter. One can imagine that Fig.~\ref{fig:puddles} shows a vertical cross section of Fig.~\ref{fig:fingers}.} 
    \label{fig:fingers}
\end{figure}

Let us start from the discussion of NNH conductivity. 
By decreasing the temperature such that $\Delta/T \gg S$ or $T \ll T_1 = \Delta /S$, NNH starts playing role and replacing the $\Delta$ activation energy by the charging energy of a puddle $E_C$.
In the case of large $\Gamma /\Delta$ we study the fractal structure of puddles which leads to a peculiar expression for $E_C$, smaller than standard 
$E_C = e^2/\kappa a$. 
Namely we are going to show that
\begin{align}\label{eq:newec}
E_C = \frac{e^2}{\kappa L} = \frac{e^2 \Delta}{\kappa a \Gamma }.
\end{align}
Substituting Eqs.~\eqref{eq:a} and \eqref{eq:lambda1} into Eq.~\eqref{eq:newec} one arrives at Eq.~\eqref{eq:ec_2}.
Let us illustrate how this happens comparing the self-capacitance of an isolated puddle  $C_{0} \sim \kappa a $ with the capacitance of the same puddle surrounded by other puddles, $C$. In the latter case, extra electron charge of a puddle $e$ is located at the distance of screening radius $\lambda$ from its border of the length $L$, while neighboring metallic puddles provides opposite charge on the other side of the border. Thus, all electric field is concentrated at the border between two puddles, mostly between long fingers of electron and hole puddles shown in Fig.~\ref{fig:fingers}. 

In our system it means that $C \sim \kappa L$ and leads to Eq.~\eqref{eq:newec}. 
The role of fingers interaction in creating large capacitance was also recognized by electrical engineering community~\cite{samavati1998}.
 
The use of the activation energy $E_C$ is justified when it is larger than the energy level spacing in a puddle.
The level spacing is given by
\begin{align}
    \delta = (gA)^{-1} = \alpha^2 (\Delta / \Gamma)^{55/36} \Delta,
\end{align}
where $g$ is the 2D DOS given by Eq.~\eqref{eq:dos} and $A$ is the area of a puddle given by Eq.~\eqref{eq:A}.
Therefore the ratio $\delta / E_C = (\Delta/\Gamma)^{7/36} \ll 1$ and our use of $E_C$ is legitimate.

Let us now switch to VRH conductivity which replaces NNH one low temperature enough temperature. 
%that $E_C/T \gg S$ or $T \ll T_2 = E_C/S$. 
In the ground state, each puddle $i$ of our system is charged by a random fractional charge $|q_{i}| \leq e/2$. 
This happens because some impurities contribute their potential to neighboring puddles effectively by sharing their charge between neighboring puddles, so that each puddle effectively gets a fraction of impurity charge $e$. 
On the other hand, electrons contribute their integer charge $e$ to their puddles. Fractional charging provides background disorder and creates random potential resulting in background density of localized states in which Coulomb attraction between excited electrons and remaining at its initial place hole produces the Coulomb gap around the Fermi level~\cite{zhang2004,chen2012}. 
This leads again to ES law in the low temperature limit.

We can calculate $T_{\text{ES}}$ in ES law starting from the standard expression $T_{\text{ES}}  = e^2/\kappa \xi$~\cite{efros1975,shklovskii1984}. Using $\xi = a/S$ 
%and $\kappa_1 = \kappa (\Gamma/\Delta)$ for $\kappa$ 
and Eq.~\eqref{eq:action_meso}
we arrive at Eq.~\eqref{eq:tes}. We see now that $T_{\text{ES}} \ll \Delta$. 
Equating $(T_{\text{ES}}/T)^{1/2}$ to $E_C/T$ with help of Eqs.~\eqref{eq:tes} and \eqref{eq:newec} we arrive at $T_2$ given by Eq.~\eqref{eq:t_2}. 

Note that at $T=T_2$ the typical hop length of ES VRH is $\xi(T_{\text{ES}}/T)^{1/2} = L\gg a$ so that the range of applicability of Eq.~\eqref{eq:newec} goes beyond the range $a$ of NNH. The reason for this is that in the case  $\Gamma \gg \Delta$ the energy of the Coulomb interaction between electron and hole at distance $r$ determining ES VRH $V(r)$ has a peculiar form: $V(r) = e^{2}/\kappa r$ at $r\gg L$ and $V(r) = e^2/\kappa L$ at $r\ll L$.\footnote{Here we would like to compare energies of two likely configurations of electric field produced by electron and hole located in the plane of a 2D semiconductor at distance $r$ from each other. First configuration corresponds to electric field lines connecting two charges through 3D space with dielectric constant $\kappa$. In this configuration the energy stored in the electric field is $e^2/\kappa r$. In the second configuration electric field stays in the plane of the semiconductor with dielectric constant $\kappa$. It runs through metallic puddles and connects their perimeters of length $\sim L$, so that puddles form capacitors with capacitance $C \sim \kappa L$. In a typical linear cross-section of the area $r^2$ there are $r/a$ parallel capacitors with charge $ea/r$ in each of them. Here $a$ is the diameter of a puddle. Each capacitor therefore carries an energy $(ea/r)^2/C$. The total number of involved capacitors is $(r/a)^2$. Thus, total energy of this capacitor network is $e^2/\kappa L$. It is clear now that at $r>L$ electric fields prefer to stay in 3D space leading to $V(r) = -e^2/\kappa r$ for electron-hole interaction energy, while at $r < L$ electric fields stay inside the plane and $V(r) = - e^2/\kappa L$ is independent on $r$. (This argument ignores logarithmic factors).}

For such a potential ES law crosses over to activated behavior Eq.~\eqref{eq:newec} when hop length $r$ becomes smaller than $L$~\cite{shklovskii2017}. Thus, NNH is responsible only for the high temperature part of the temperature range of validity of Eq.~\eqref{eq:newec}, the second low temperature one is ES VRH corresponding to $V(r) = e^2/\kappa L$.

Above we studied the simplest case of the gapped Dirac spectrum Eq.~\eqref{eq:spectrum}. 
In an important case of BLG gapped by perpendicular external electric field~\footnote{Graphene has a Fermi velocity $v=1\times 10^{6}$ m/s, and in order to apply our theory we need $\alpha=e^2/\kappa \hbar v \simeq 2.2/\kappa \ll 1$. Namely, the dielectric constant of the environment surrounding the BLG should be $\kappa \gg 2.2$.}, the spectrum is somewhat different, namely, it has the ``mexican hat'' shape, where the energy minimum $|\epsilon(\vb{k})|=\Delta$ is degenerate and located along a ring $\abs{k} = k_0$ in 2D $\vb{k}$-space~\cite{mccann2013}. 
Nevertheless, at $|\epsilon(\vb{k})|\gg \Delta$ the spectrum returns to the same Dirac cone as Eq.~\eqref{eq:spectrum} and $|\epsilon(\vb{k})| \sim \Delta$ is the only characteristic low energy scale. 
This is why Eqs.~\eqref{eq:gamma} and ~\eqref{eq:lambda} are still valid and the order of magnitude of $S$ is not changed. 
Thus, all our results for the most interesting case $\Gamma \gg \Delta$ are still the same as for the gapped Dirac spectrum Eq.~\eqref{eq:spectrum}. 

\section{Modest concentration of impurities}
\label{sec:lowdisorder}

Above we assumed the gap $\Gamma \gg\Delta$ so that we used the Dirac dispersion $\epsilon(k) \approx \hbar v k$ to calculate energies $E_C$ and $T_{\rm ES}$.
In this section we study the opposite case when $\Gamma \ll \Delta$ or $N \ll N_0$, where $N_0$ is given by  Eq.~\eqref{eq:N0}. 
In this case, $\Gamma$ given by Eq.~\eqref{eq:gamma} does not describe the potential fluctuation amplitude and $\Gamma$ can be considered only as a measure of $N$. Indeed, when the Fermi level is within the gap there is no screening, unless the Coulomb potential bends the conduction band bottom and the valence band ceiling by an energy slightly larger than $\Delta$ and creates small electron and hole droplets. Electron and hole droplets form alternating in space fractal electron and hole clouds (puddles) of the size
\begin{align}\label{eq:big_lambda}
    \Lambda = \kappa^2 \Delta^2 / e^4 N,
\end{align}
obtained by equating $\Delta$ and the random potential amplitude $(e^2/\kappa)(N\Lambda^3)^{1/2} /\Lambda$ inside a cube of size $\Lambda$. 
Such a system of droplets and puddles is the two-dimensional analog of the three-dimensional completely compensated semiconductor~\cite{shklovskii1972,shklovskii1984} schematically shown in Fig.~\ref{fig:large_gap}. 

If the kinetic energy of degenerate electrons in droplets satisfies $\epsilon(k) \ll \Delta$, then one can use the parabolic dispersion law for them
\begin{align}\label{eq:dispersion_nr}
    \epsilon(k) = \hbar^2 k^2/2m,
\end{align}
with $m = \Delta/v^2$. To show that indeed $\epsilon(k) \ll \Delta$  we first find the size of a typical droplet $R_q$, following Refs.~\cite{shklovskii1972,shklovskii1984}. Namely, we equate the depth of the potential well $(e^2/\kappa)(NR_q^3)^{1/2}/R_q$ created by a typical fluctuations of charge in a cube of size $R_q$ to the kinetic energy of $(NR_q^3)^{1/2}$ electrons $\epsilon =(NR_q^3)^{1/2} \hbar^2/mR_q^2$ in the disk of radius $R_q$ and arrive at $R_q = a_B$, where $a_B = \hbar^2 \kappa/ me^2$ is the semiconductor Bohr radius and $m=\Delta/v^2$ is the effective mass~\footnote{The same result can be obtained by equating the total number of electron states $g \gamma(R_q) R_q^2$ to the excess number of impurity charges $\sqrt{NR_q^3}$, where $\gamma(R_q)=(e^2/\kappa)(NR_q^3)^{1/2}/R_q$ is the potential depth and $g = m/\hbar^2$ is the 2D DOS.}. When deriving $R_q=a_B$ we assumed $N a_B^3 = (\Gamma/\Delta)^3 \alpha^{-4} \gg 1$, or equivalently $1 \gg \Gamma/\Delta \gg \alpha^{4/3}$.
Substituting $R_q=a_B$ back to $\epsilon =\hbar^2 (NR_q^3)^{1/2}/mR_q^2$ we get
\begin{align}\label{eq:ratio}
    \frac{\epsilon}{\Delta} = \qty(\frac{e^2 N^{1/3}}{\alpha^{2/3}\kappa \Delta})^{3/2} = \qty(\frac{\Gamma}{\Delta})^{3/2} =\qty(\frac{N}{N_0})^{1/2} \ll 1.
\end{align}

\begin{figure}[t]
    \centering
    \includegraphics[width=\linewidth]{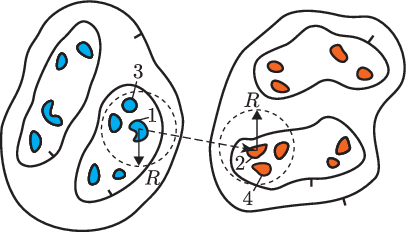}
    \caption{Schematic map of nearest neighbor electron and hole puddles containing many electron (blue) and hole (red) puddles. The continuous lines are equipotential contours of the electron energy. As in geographical maps the direction of descent is indicated by a short stroke. The smallest contours represent boundaries of droplets at the chemical potential. The dashed arrow shows the shortest hop between the two puddles. At $T\ll T_1'$ electron searches in dashed circles of radius $R$ for droplets 3 and 4 with closer to the chemical potential energies, which provide a smaller inter-puddle hop resistance.} 
    \label{fig:map}
\end{figure}

Next we discuss the conductivity when $N \ll N_0$ or $\Gamma \ll \Delta$.
Similarly to the case $\Gamma/\Delta \gg 1 $ there are three mechanisms of the conductivity.
Activation of free electrons by energy $\Delta$ is the same as at $N> N_0$. 
The tunneling between droplets inside the same puddle is faster than the tunneling between neighbor puddles. 
For the latter we can find the action $S$ substituting $\Lambda$ for $r$ into Eq.~\eqref{eq:action}:
\begin{align}\label{eq:action_lambda}
    S = \alpha^{-1} (\Delta/\Gamma)^3 = \alpha^{-1} N_0/N.
\end{align}
This leads to the crossover from activation to hopping at $T_1' = \Delta/S$ or
\begin{align} \label{eq:t_1p}
T_1' = \alpha (N/N_0) \Delta.
\end{align}
In the adjacent interval of lower temperatures $T_1' \gg  T \gg T_2'$ where 
\begin{align} \label{eq:t_2p}
T_2' = \alpha^{3}(N/N_0)^{2}\Delta,
\end{align}
we deal with the hopping conductivity of electrons between nearest puddles, but each puddle has many droplets and we should explore hopping between all pairs of droplets of two neighboring puddles searching for the smallest Miller-Abrahams resistors~\cite{shklovskii1984}. Thus, focusing on puddles we deal with NNH and focusing on droplets we deal with VRH. Therefore, we call this mechanism of the hybrid hopping conductivity and illustrate it in Fig.~\ref{fig:map}. At $T_3 \ll T \ll T_1'$, where as we show below
\begin{align} \label{eq:t_3}
T_3 =\alpha^3 (N_0/N)^{1/4} \Delta \gg T_2'
\end{align}
an electron hops between two closest droplets 1 and 2 of the nearest neighbor electron and hole puddles. 
When temperature gets lower it chooses slightly more distant droplets of the same two puddles with energies closer to the Fermi level. At low enough temperature, $T=T_2'$, it chooses the optimal in energy two droplets of the two puddles. At even lower $T < T_2'$ it hops to the second nearest puddle and we arrive at ES VRH. Such a theory of  hybrid hopping conductivity was proposed for strongly compensated semiconductors~\cite{shklovskii1973} before the discovery of the Coulomb gap and ES law~\cite{efros1975}. Here we revise this theory by adding the Coulomb gap to its argumentation.

To find the optimal path at $T_2'\ll T \ll T_3$, an electron searches for more distant distant droplets 3 and 4 located in two discs of radius $R$ around droplets 1 and 2. This leads to the increase $\delta S = R/\xi(R)$ of the tunneling action $S$, where $\xi(R) =\hbar/(m\gamma(R))^{1/2}$ and $\gamma(R)= (e^2/\kappa R) (NR^3)^{1/2}$ is height of typical barrier in the spacial scale  $R$. This increase can be overcompensated by the reduction of activation energy necessary for the hop between new pair of droplets. Using the two-dimensional Coulomb gap density of states $g(\epsilon) =\kappa^2 \epsilon/e^4$ we find that minimum energy $\epsilon_{\min}$ of a droplet in a disc of area $R^2$ is determined by the condition $\epsilon = [g(\epsilon)R^2]^{-1}$, which leads to $\epsilon_{\min} = e^2/\kappa R$. 

Thus, the logarithm of conductivity generated by optimization in a given scale $R$ is 
\begin{align} \label{eq:sigmar}
\ln\sigma(R) = -S - R [m\gamma(R)]^{1/2}/\hbar - e^2/\kappa R T.
\end{align}
It has maximum at
\begin{align}\label{eq:roptimal}
R_{\rm opt} = e^2 \kappa^{-1} T_{H}^{-5/9}T^{-4/9},
\end{align}
where $T_H = \alpha^{6/5}(N/N_0)^{1/5}\Delta$ as shown in Eq.~\eqref{eq:t_H}.
Substituting $R_{\rm opt}$ back to Eq.~\eqref{eq:sigmar} we arrive at Eq.~\eqref{eq:sigma_hybrid}.
Equating $R_{\rm opt}$ to the maximum distance $\Lambda$ between droplets inside a puddle we arrive at $T_2'$ in Eq.~\eqref{eq:t_2p}. 
On other hand, equating $R_{\rm opt}$ to the minimum distance $a_B$ between droplets inside a puddle we arrive at the high temperature limit of the hybrid hopping conductivity $T_3$ given by Eq.~\eqref{eq:t_3}.

When $N$ tends to $N_0$ the temperature $T_3$ tends to $T_2'$, so that the range of validity of Eq.~\eqref{eq:sigma_hybrid} vanishes.
This allows the conductivity we calculated Sec.~\ref{sec:lowdisorder} for $N > N_0$ to match at $N=N_0$ the conductivity we found for $N < N_0$ in Sec.~\ref{sec:hopping}.~\footnote{Above for simplicity we concentrated on the case of $N > \alpha^{8/5} N_0$, when $T_3 > T_1'$. In this case, in the range $T_1' > T > T_3$ conductivity has a constant activation energy. At $N < \alpha^{8/5} N_0$ the high temperature border of the hybrid hopping conductivity Eq.~\eqref{eq:sigma_hybrid} is given by $T_1'$}

At lower temperatures $T < T_2'$, the conductivity is dominated by ES VRH.
The localization length is given by 
\begin{align}\label{eq:xi_d}
    \xi = \Lambda/S,
\end{align}
with $\Lambda$ given by Eq.~\eqref{eq:big_lambda} and $S$ given by Eq.~\eqref{eq:action_lambda}.
Substituting Eq.~\eqref{eq:xi_d} into $T_{\rm ES} = e^2/\kappa \xi$, we get the characteristic temperature for ES VRH
\begin{align}\label{eq:tes_d}
    T_{\rm ES} = \alpha \Delta,
\end{align}
which is valid if $\alpha^4 < N/N_0 <1 $ and matches Eq.~\eqref{eq:tes} at $\Gamma = \Delta$ and $N=N_0$. 

Our results for $T_{\rm ES}$ and $E_C$ as a function of dimensionless impurity concentration $N/N_0 = (\Gamma/\Delta)^3$ obtained in Sections \ref{sec:hopping} and \ref{sec:lowdisorder} are summarized in Fig.~\ref{fig:energy_n}. 
%Dramatic collapse of $T_{\rm ES}$ at $N > N_0$ is the main result of this theory.

So far we assumed that $\alpha \ll 1$. Let us explore what happens when $\alpha \sim 1$ as, for example, in twisted bilayer graphene. 
In this case at $N_0 = \kappa^3\Delta^{3}e^{-6} = a _B^{-3}$ and at $N=N_0$ we have $Na_B^3 = 1$. Both sides of $N/N_0 =1$  on Fig.~\ref{fig:energy_n} now shrink. At $N/N_0 \gg 1$ or $Na_B^3\gg 1$ we get almost metallic conductivity, in spite of complete compensation. On the other side, $N/N_0 \ll 1$, 
the conductivity obeys ES law with $T_{\rm ES} = \Delta$ at low temperature $T \ll (N/N_0)^{2}\Delta$. At higher temperatures interval $(N/N_0)^{2}\Delta \ll T \ll  (N/N_0)\Delta$ the conductivity is due to the hybrid hopping described by Eq.~\eqref{eq:sigma_hybrid} with $T_H = \Delta (Na_B^3)^{1/5}$. At $T \gg (N/N_0)\Delta$ the conductivity is dominated by free electrons activated by energy $\Delta$. As a result, at low enough temperature the local apparent activation energy is of the order of $(T \Delta)^{1/2} \ll \Delta$. Note also that at $\alpha \sim 1$ the Bohr energy of hydrogen like impurities $E_B=e^2/\kappa a_B=\alpha^2 \Delta$ becomes of the order of $\Delta$ so that localized donor and acceptor states play the role of metallic droplets in the lightly doped limit $N \ll N_0$ (see Chapter 3 of Ref.~\onlinecite{shklovskii1984}).

\begin{figure}[t]
    \centering
    \includegraphics[width=\linewidth]{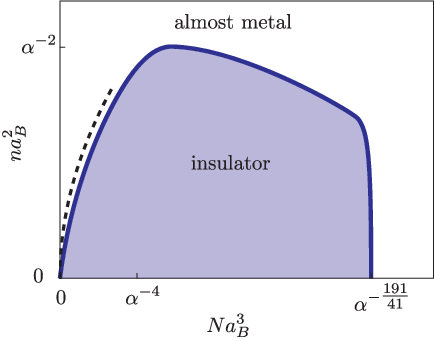}
    \caption{Schematic $n$-$N$ phase diagram in a two-dimensional semiconductor. The shaded blue domain is the insulator phase while the white domain is the almost metal phase. On the left (small $N$) side of the diagram the phase boundary follows Eq.~\eqref{eq:n_c} (dashed line) and reaches the maximum near $N a_B^3 = N_0 a_B^3 = \alpha^{-4}$. On the right side the maximum of the phase boundary is determined by criterion $G(n,N)=1$ for tunneling between electron puddles. When with decreasing $n$ this tunneling rate yields to the tunneling between electron and hole puddles, the boundary becomes vertical, i. e., sticks to the critical point $N a_B^3 = N_c a_B^3 = \alpha^{-191/41}$ all the way till $n=0$ [c.f. Eq.~\eqref{eq:critical}]. We use $\alpha = 0.12$ for this plot.} 
    \label{fig:phase1}
\end{figure}

\section{``Phase diagram'' of transition between insulator and almost metal}
\label{sec:PD} 

Above we focused on the charge neutrality point where $E_F = 0$. However, for 2D devices one can easily move away from the neutrality point by applying a gate voltage, for example, making $E_F > 0$ and inducing a non-zero net 2D concentration $n$ of electrons.
As a result one can study the whole $n$ -- $N$ phase diagram. We show below that such a phase diagram has an interesting re-entrant metal-insulator transition (MIT) as a function of increasing $N$ (see Fig.~\ref{fig:phase1}). 

\begin{figure}[t]
    \centering
    \includegraphics[width=\linewidth]{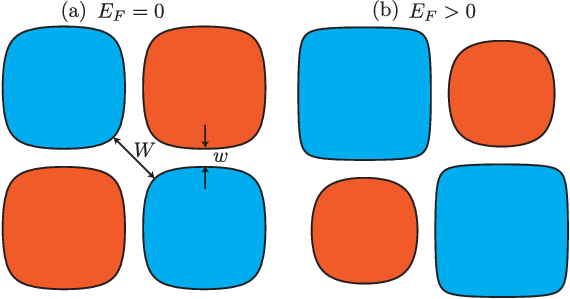}
    \caption{Illustration of the competition of the two tunneling rates for a chessboard potential. Blue and red domains are the electron and hole puddles separated by insulating gap (white). a) At the neutrality point, $n=0$, $E_F = 0$, the shortest tunneling distance between electron puddles $W$ is much larger than the distance $w$ between electron and hole puddles. b) At $E_F > 0$ and growing $n$, $W$ decreases. Eventually it becomes smaller than $w$ and vanishes at the percolation transition where all electron puddles merge into the infinite cluster.} 
    \label{fig:fourpuddles}
\end{figure}

We already know from Eq.~\eqref{eq:critical} that if $n=0$, there is an insulator-metal transition (IMT) at $\Gamma/\Delta = (\Gamma/\Delta)_c$, or $N= N_c=N_0\alpha^{-27/41}$, where the disorder tunneling  plays decisive role.
On the other hand, for modest impurity concentrations $1 \ll Na_B^3 \ll N_0 a_B^3 = \alpha^{-4}$, there is also an MIT induced at some $n=n_c$. This MIT happens because by increasing $n$ the electron kinetic energy becomes larger than the random Coulomb potential energy, and electron puddles become connected with each other and turn the system from an insulator to an almost metal. We find the net electron concentration $n_c$ associated with the MIT by equating the Fermi energy $E_F = \hbar^2 n/m$ with the Coulomb potential energy fluctuations~\cite{huang2021b}
\begin{align}\label{eq:comp_eq}
    \frac{\hbar^2 n}{m} = \frac{e^2\sqrt{N a_B^3}}{\kappa a_B},
\end{align}
where the Bohr radius $a_B$ plays the role of the linear screening length $r_s$ for non-relativistic 2D electron gas. 
Solving Eq.~\eqref{eq:comp_eq} one gets the critical percolation threshold concentration
\begin{align}\label{eq:n_c}
    n = n_c = (N/a_B)^{1/2}.
\end{align}
In the left side of the phase diagram Fig.~\ref{fig:phase1}, the MIT border line follows Eq.~\eqref{eq:n_c} shown by the dash line.
It continues till the maximum at $N\sim N_0$, where $E_F \sim \Gamma \sim \Delta$. Therefore, for the whole phase diagram $E_F \leq \Delta$ and our use of the non-relativistic expressions for $E_F$ and $r_s$ in Eq.~\eqref{eq:comp_eq} is justified. 

Now we turn to the right side of Fig.~\ref{fig:phase1} and find the IMT border line, which starts at $n=0$ and $N= N_c$ and matches Eq.~\eqref{eq:n_c} near $N= N_0$. 
Apparently, this part of the border is determined by tunneling between puddles. 
%So far we dealt with the tunneling between electron and hole puddles. 
This tunneling can be illustrated by the small square section of the ``chess-board'' potential $u(x,y) = u_0 \cos(2\pi x/b)\cos(2\pi y/b)$ which at neutrality point has four identical electron and hole puddles (see the $b \times b$ square in Fig.~\ref{fig:fourpuddles}). There are two kinds of electron tunneling events contributing in the conductivity of such a chess-board, the side-to-side tunneling between electron and hole puddles and the diagonal tunneling through the saddle point between two neighboring electron puddles. 

At $n=0$ [see Fig.~\ref{fig:fourpuddles} (a)] in both cases the tunneling barrier is of the same height $\Delta$, while the tunneling distance in the side-to side case, $w$ is substantially shorter than than the diagonal distance $W$. Thus, at $n=0$ the side-to-side tunneling dominates and so far in this paper we dealt only with this tunneling. The side-to-side tunneling distance $w$ is weakly affected by growing $n$ [see Fig.~\ref{fig:fourpuddles} (b)]. As a result the right side of the IMT border at small $n$ is almost a vertical line. On the other hand, we see in Fig.~\ref{fig:fourpuddles} (b) that the distance between electron puddles $W$ steeply decreases with growing $n$. Eventually, $W$ becomes smaller than $w$ and the diagonal tunneling starts dominating.

Although in Fig.~\ref{fig:fourpuddles} we used a simplified model, the same dynamics is valid for our disordered system. The crossover to the diagonal tunneling leads to a sharp decrease of the slope of the right side border of the insulating phase of Fig.~\ref{fig:phase1},
given by the condition  $G(n,N) = 1$ for the dimensional conductance. Eventually, with growing $n$ large fraction of diagonal gaps between electron puddles close near $na_B^2 \sim \alpha^{-2}$, where $E_F \sim \Delta \sim \Gamma$ and electron puddles merge at the percolation transition.
This means that the right side border merges with the left one near $N \sim N_0$ as shown in Fig.~\ref{fig:phase1}. 
 
We see that at $1 < n a_B^2 < \alpha^{-2}$ there is an unconventional re-entrant MIT. At small $N \ll N_0$ disorder drives a percolation MIT, when $\Gamma$ exceeds Fermi energy of electrons so that they are forced into isolated puddles. On other hand, at large $N > N_0$ or $\Gamma > \Delta$ the disorder drives electrons back to metal, because of increase of tunneling which becomes dramatic near $ \Gamma/\Delta = (\Gamma/\Delta)_c$. Of course, making $n < 0$ one can explore re-entrant MIT phase diagram for holes.

Changing the concentration of impurities $N$ at fixed $n$ and $\Delta$ in real experiments is difficult. 
Instead, one can change the band gap $\Delta$ at fixed $n$ and $N$ as this is done, for example, in Ref.~\onlinecite{Li2021}. However, in this case, with decreasing $\Delta$ there is only one MIT at small $\Delta$ as in the case of $n=0$. There is no second transition at large $\Delta$, because as we see from Eq.~\eqref{eq:n_c} 
using $a_B = e^2/(\kappa\alpha^2\Delta)$, the critical concentration of MIT $n_c\propto \Delta^{1/2}$ eventually becomes larger than any fixed $n$, so that the insulator phase persists. One can return to a re-entrant MIT transition at fixed $N$ by changing $\Delta$ and $n$ simultaneously, for example, by fixing $n a_B^2$ as this is clear from Fig.~\ref{fig:phase1}.

We would like to emphasize that predicted in this section reentrant MIM transition is different from MIM trasnition observed near filling factor $\nu =1$, where the Mott-Hubbard gap has maximum~\cite{Ghiotto2021}. Along a horizontal line of Fig.~\ref{fig:phase1}, $\Delta$ and $n$ are fixed and the concentration of impurities $N$ changes monotonically.
Monotonically growing random potential then leads to the classical MIT via creating puddles when $N$ is relatively small, and to the quantum IMT transition via enhanced tunneling when $N$ is big. 
For the transition discussed in our previous paragraph, $N$ is fixed
and both $n$ and $\Delta$ change monotonically, such that two transitions are due to the interplay of two different, classical and quantum, disorder effects.
Our MIM transition exists when there is no Mott-Hubbard physics.

\section{Thin film of three-dimensional topological insulator}
\label{sec:TI}

In previous sections we dealt with the general model of the trivial 2D semiconductor with gaped Dirac spectrum Eq.~\eqref{eq:spectrum}. 
In this section we concentrate on special case of a thin film of 3D TI, where the narrow gap $2\Delta$ can be a result of the hybridization of surface states on opposite surfaces of the film~\cite{liu2010,lu2010,linder2009,zhang2010a,sakamoto2010,zhang2013,kim2013,nandi2018,chong2021} or created by a concentration of magnetic dopants like Cr~\cite{chen2010,yu2010,xu2012,checkelsky2012,jinsongzhang2013,he2013,mogi2015,ye2015,liguozhang2017,wang2018,fox2018,moon2019,rosen2019,tokura2019,okazaki2020,deng2020,rodenbach2021,lu2021,fijalkowski2021,ferguson2021,rosen2021,okazaki2022}. 
Because of the promise of such films to achieve a metrological precision of the quantum anomalous Hall effect and the quantum spin Hall effect,
in this section we are more specific with material parameters and numerical estimates.

We have in mind TI thin films based on (Bi$_{x}$ Sb$_{1-x}$)$_2$Te$_{3}$, which have very large dielectric constant $\kappa \sim 200$~\cite{richter1977,borgwardt2016,bomerich2017}. 
Using $\kappa \sim 200$ and the Fermi velocity of TI $v \sim 4 \times 10^5$ m/s~\cite{jszhang2011}, one gets $\alpha \sim 0.027$.
We assume that such a film of width $d \sim 7$ nm is deposited at the substrate with much smaller dielectric constant $\kappa_e$, so that electric fields of Coulomb impurities residing inside the film are trapped within the film~\cite{rytova1967,chaplik1971,keldysh1979,huang2021a,huang2021c}. 
Namely, electric lines stay inside the film at distances $r$ from impurity if $d< r< d\kappa/\kappa_e$ and in the absence of screening exit the film at $r > d\kappa/\kappa_e$. Thus, the electric field of a charged impurity is $E(r) = 2e^2/\kappa dr$ at $d< r< d\kappa/\kappa_e$ and $E(r) = e^2/\kappa_e r^2$ at $r> d\kappa/\kappa_e$. This leads to electrostatic potential $V(r) = (e^2/\kappa d) \ln(\kappa d/r)$ weakly dependent on $r$ in the range $d< r< d\kappa/\kappa_e$.

At $\Gamma > \Delta$ electrons and holes created by band bending self consistently screen electric field at the distance $\lambda$, such that $d \ll\lambda\ll (\kappa/\kappa_e)d$. Very slow decay of the electrostatic potential of distant Coulomb impurities allows larger number of them to contribute to $\Gamma$. TI films have large concentrations of Coulomb impurities, $N \sim 10^{19}$ cm$^{-3}$, so that we study only their effect and ignore impurities inside the substrate. 

Peculiar electrostatics we discussed above leads to new results of self-consistent theory of screening~\cite{huang2021a} at $E_F=0$ and $\Gamma \gg \Delta$ 
\begin{align}\label{eq:gamma_ti}
    \Gamma & = \frac{e^2 N^{1/3}}{\kappa \alpha^{2/3}}, \\
    \lambda & = \alpha^{-2/3} (Nd^3)^{-1/6} d.\label{eq:lambda_ti}
\end{align}

Note that the expression for $\Gamma$ in this case is the same as Eq.~\eqref{eq:gamma}, i.e. while the expression for $\lambda$ differs from Eq.~\eqref{eq:lambda} and is valid if $\lambda \gg d$. Using above estimates for $\alpha$, $d$ and $N$ we get $\lambda \sim 50$ nm, so that inequality $\lambda \gg d$ holds and we can use Eq.~\eqref{eq:lambda_ti}.

Similarly to the Sec.~\ref{sec:hopping}, below we calculate the tunneling action $S$ and the critical $(\Gamma/\Delta)_c$ and describe hopping conductivity of the film. 
Notice the electric field in the film plane created by charge fluctuations in a disk of radius $r$ and thickness $d$ is given by
\begin{align}\label{eq:E_ti}
    E = \frac{e \sqrt{Nr^2 d}}{\kappa r d} = \frac{e}{\kappa} \sqrt{\frac{N}{d}},
\end{align}
which turns out to be independent on $r$.
Therefore, there is no, similar to the one in Sec. IV, enhancement of the electric field at scales shorter than $\lambda$. 
Substituting Eq.~\eqref{eq:E_ti} or equivalently $E=\Gamma/e\lambda$ into Eq.~\eqref{eq:action}, we arrive at the action
\begin{align}\label{eq:action_w_ti}
    S = \frac{w \Delta}{\hbar v} = \frac{\lambda  \Delta^2}{\hbar v \Gamma} =  \alpha^{-1/3} (Nd^3)^{1/6} (\Delta / \Gamma)^{2}.
\end{align}
However, the electric field $E = eM/\kappa \lambda d$ can still be enhanced by a rare fluctuation of the number of charges $M \gg (N\lambda^2 d)^{1/2}$ with
Gaussian probability $\exp(-M^2/N\lambda^2 d)$. This replaces Eq.~\eqref{eq:m} by
\begin{align}
    (L/\lambda) \exp[-\frac{M^2}{N\lambda^2d}] = 1.
\end{align}
Solving the above equation we obtain the largest $M$ available in the perimeter
\begin{align}
    M = \qty{N\lambda^2 d \ln[(\Gamma/\Delta)^{7/3}]}^{1/2}.
\end{align}
Substituting the electric field $E = eM/\kappa \lambda d$ into the action Eq.~\eqref{eq:action} we obtain
\begin{align}\label{eq:action_meso_ti}
    S = \frac{\alpha^{-1/3} (N d^3)^{1/6} (\Delta / \Gamma)^{2}}{\qty{\ln[(\Gamma/\Delta)^{7/3}]}^{1/2}} \simeq \alpha^{-1/3} (N d^3)^{1/6} (\Delta / \Gamma)^{43/18},
\end{align}
which is smaller than the action given by Eq.~\eqref{eq:action_w_ti}. In the last step, as in the Sec.~\ref{sec:fractal} we used the power-law approximation $\ln x=x^{1/3}$ valid for $x \in (3,100)$ with accuracy better than 15\%.

Substituting Eq.~\eqref{eq:action_meso_ti} into the expression of $G$, Eq.~\eqref{eq:g}, and setting $G=1$, we arrive at the critical point~\footnote{In the limit of $\alpha (N d^3)^{-1/2}\to 0$, the asymptotic expression to first order reads $(\Gamma/\Delta)_c = \alpha^{-1/6} (N d^3)^{1/12}\qty{\ln[\alpha^{-1} (N d^3)^{1/2}]}^{-3/4}$.}
\begin{align} \label{eq:critical_ti} 
    (\Gamma/\Delta)_c = \alpha^{-2/19}(N d^3)^{1/19}.
\end{align}
Using the estimates $\alpha \sim 0.027$, $N \sim 10^{19}$ cm$^{-3}$ and $d=7$ nm we get $(\Gamma/\Delta)_c = 1.6$.

Let us switch to hopping conductivity of thin TI film and start from NNH where activation energy is determined by the charging energy of a puddle.
Similarly to Sec.~\ref{sec:fractal}, capacitance of a puddle within the film is determined by its long border adjacent to neighboring puddles.
Near the border there are two stripes with charges $-e$ and $e$ of the length $L$ and width $\lambda$. 
But now electric field at the border is concentrated in the film of width $d\ll\lambda$ with the large dielectric constant $\kappa$. This changes the capacitance of the puddle border to $C\sim \kappa L(d/\lambda)$ and leads to 
\begin{align}\label{eq:ec_ti}
    E_C =  (e^2/\kappa d)(\lambda/L) = (e^2/\kappa d)(\Delta / \Gamma)^{7/3},
\end{align}
%\begin{align}\label{eq:ec_ti}
    %E_C = \alpha^{4/3} (Nd^3)^{-1/6} (\Delta / \Gamma)^{4/3} \Delta,
%\end{align}
at $1 < \Gamma/\Delta < (\Gamma/\Delta)_c$. This charging energy, plays the role of NNH activation energy. 

At lower temperatures $T<T_2$ conductivity obeys the ES law with the characteristic temperature $T_{\rm ES} = e^2/\kappa_e \xi$, where $\xi = a/S$. 
Note that here we use $\kappa_e$ because at large distances electric field lines leave the film and go through the environment. Using Eqs.~\eqref{eq:a} and \eqref{eq:action_meso_ti} we get
\begin{align} \label{eq:tes_ti}
    T_{\rm ES} = \alpha (\kappa/\kappa_e) \Delta (\Delta/\Gamma)^{49/18}.
\end{align}
Crossover between Eq.~\eqref{eq:sigmaD} and  Eq.~\eqref{eq:sigma_NNH} happens at
\begin{align} \label{eq:t_4}
    T_4 =  \alpha^{1/3}(Nd^3)^{-1/6}(\Gamma/\Delta)^{43/18}\Delta,
\end{align}
while crossover between Eq.~\eqref{eq:sigma_NNH} and  Eq.~\eqref{eq:sigma_ES} happens at 
\begin{align} \label{eq:t_5}
    T_5 = E_c^2/T_{\rm ES} = \alpha^{-1}(\kappa_e/\kappa)(\Delta/\Gamma)^{35/18} \Delta^{-1} (e^2/\kappa d)^2
\end{align}    

Eqs.~\eqref{eq:tes_ti} and ~\eqref{eq:critical_ti} show that in TI films as in trivial semiconductors [c.f. Eqs.~\eqref{eq:tes} and ~\eqref{eq:critical} and Fig.~\ref{fig:energy_n}] reduction of $T_{\rm ES}$ and crossover from strong localization case to the practically metallic conductivity happens dramatically fast when $\Gamma$ exceeds $\Delta$.

Now let us study the case of a modest concentration of impurities where $\Gamma < \Delta $ or $N < N_0$ where $N_0$ is defined by Eq.~\eqref{eq:N0}.
First we should find the size of a droplet $R_q$ in the cloud, and show that the kinetic energy of a droplet $\epsilon$ is smaller than $\Delta$ so that the energy dispersion is non-relativistic at $\Gamma < \Delta$. 
Equating the depth of the potential well $e^2(NR_q^2d)^{1/2}/\kappa d$ created by typical fluctuations of charges in a disk of radius $R_q$ and thickness $d$ to the Fermi energy of $(NR_q^2d)^{1/2}$ electrons in the disc of radius $R_q$, $\epsilon =\hbar^2 (NR_q^2d)^{1/2}/mR_q^2$ , we arrive at $R_q = \sqrt{a_B d}$.
This $R_q$ is valid if $a_B > d$ (or $\Delta < \alpha^{-2} e^2/\kappa d$) so that the droplet is still disk-like.
Substituting $R_q = \sqrt{a_B d}$ back to the droplet Fermi energy we get the ratio $\epsilon / \Delta$ given by Eq.~\eqref{eq:ratio}.
Therefore, at  $N < N_0$  the kinetic energy is smaller than $\Delta$ and the use the parabolic dispersion law is justified. 

Next, we study the three-mechanism conductivity at $N < N_0$.
Let us start with the NNH conductivity.
Similarly to Sec.~\ref{sec:hopping}, we obtain the nonlinear screening length 
\begin{align} \label{lambda_ti2}
    \Lambda = \frac{\Delta \kappa \sqrt{d}}{e^2 \sqrt{N}}.
\end{align}
(the size of a puddle) by equating $\Delta$ and the random potential of a disk of size $\Lambda$.
\begin{align}
    \Delta = \frac{e^2}{\kappa d} \sqrt{N d \Lambda^2}
\end{align}
As a result, the tunneling action $S$ is again given by Eq.~\eqref{eq:action_w_ti} and $S \gg 1$.
The activation energy for the NNH conductivity is equal to the charging energy of a droplet. 
Radial electric field of a charged droplet stays in insulating TI film till distance $(\kappa/\kappa_e)d$ 
and only then exits to the environment with small dielectric constant $\kappa_e$.
Therefore, the charging energy determining hopping activation energy at $T_4 > T > T_5$ is
\begin{align}\label{eq:ec_ti_d}
    E_C = e^2/[\kappa_e (\kappa/\kappa_e) d] =\frac{e^2}{\kappa d},
\end{align}
which matches Eq.~\eqref{eq:ec_ti} at $\Gamma = \Delta$\footnote{This result of $E_C$ is valid if the size of a droplet $R_q$ is smaller than the electric field confinement distance $(\kappa/\kappa_e) d$, or $\Delta > (\kappa_e/\kappa \alpha)^2 e^2/\kappa d$.}.

At lower temperature $T < T_5$, using the localization length $\xi = \Lambda/S$ we get the characteristic temperature for ES law 
\begin{align}\label{eq:tes_ti_d}
T_{\rm ES} = e^2/\kappa_e \xi = \alpha(\kappa/\kappa_e) \Delta, 
\end{align}
which matches Eq.~\eqref{eq:tes_ti} at $\Gamma = \Delta$~\footnote{The above theory assumes that $\Lambda < d\kappa/\kappa_e$, or $\Delta < \Delta_1 = (e^2/\kappa_e) \sqrt{Nd}$. In this case, the Coulomb interaction energy between two droplets inside a puddle is equal to $e^2/\kappa d$ and is independent on the distance between two droplets $R$. Therefore, there is no optimization with respect to $R$ similar to Eq.~\eqref{eq:sigmar}, and no hybrid mechanism. On the other hand if $\Delta > \Delta_1 $, there is intermediate between Eqs.~\eqref{eq:ec_ti_d} and ~\eqref{eq:tes_ti_d} hybrid regime which we skip here.}.

Now we can estimate the characteristic energies $\Gamma$, $\Delta$, $E_C$, $T_{\rm ES}$, $T_4$, $T_5$ and $\Delta_1$ for TI thin films based on (Bi$_{x}$ Sb$_{1-x}$)$_2$ (Te$_{y}$ Se$_{1-y}$)$_3$. Using $\kappa=200$, $\alpha = 0.027$, $N = 10^{19}$ cm$^{-3}$, we have $\Gamma \simeq 17$ meV. 
The hybridization gap is related to the thickness by $\Delta = \Delta_0 e^{-d/d_0}$ with $\Delta_0 = 0.5$ eV and $d_0=2$ nm~\cite{chong2021}. 
For example, if $d = 7$ nm, then $\Gamma \simeq 17$ meV, $\Delta \simeq 15$ meV, $S \simeq 3$, $E_C \simeq 0.8$ meV, $T_4 = \Delta/S \simeq 60$ K,  $T_5 = E_C^2/T_{\rm ES} = 0.6$ K, $T_{\rm ES} \simeq 130$ K and $\Delta_1 = 70$ meV (here assume that the film has hBN on both sides and use $\kappa_e =5$). In this case, ES conductivity starts when $(T_{\rm ES}/T_5)^{1/2} \sim 15$, so large that ES law is hardly observable because of very large resistance. Thus, observable activation energy is given by $E_C \sim 0.05 \Delta$. 

In slightly thicker films with $d\geq 8$ nm the half-gap $\Delta(d) \leq 9$ meV and $\Gamma/\Delta > (\Gamma/\Delta)_c$, so that they are almost metallic and show ES conductivity with much smaller $T_{\rm ES}$. On the other hand, in slightly thinner films, $d < 7$ nm, for which $\Delta > \Gamma $ energies $E_c$ and $T_{\rm ES}$ given by Eqs.~\eqref{eq:ec_ti_d} and ~\eqref{eq:tes_ti_d} match Eq.~\eqref{eq:ec_ti} and ~\eqref{eq:tes_ti} so that practically conductivity is similar to films with $d = 7$ nm.

Notice that critical thickness $d=d_c =7$ nm is very sensitive to values of 
$\Delta_0$, $N$, $\kappa$, $\alpha$, and most importantly $d_0$, which are different for different materials. 
This can explain differences between experimental results in Refs.~\onlinecite{nandi2018,chong2021}.

For the case of magnetically doped TI thin films, the exchange half gap $\Delta$ induced by magnetic impurities is of order of 20 meV~\cite{rodenbach2021,lu2021,fijalkowski2021} (which is not directly related to $d$), so that we have practically the same numbers as in the previous example.

Let us now dwell on the $n$-$N$ plane phase diagram of the TI film, which as we show below looks almost identical to Fig.~\ref{fig:phase1}.  
At a modest impurity concentration $N < N_0$, we equate the Fermi energy with the Coulomb potential energy fluctuations to find the critical concentration $n_c$
\begin{align}\label{eq:comp_eq_ti}
    \hbar^2 n/m =  e^2\sqrt{N r_s^2 d}/\kappa d = e^2\sqrt{N a_B}/\kappa,
\end{align}
where we use $r_s=\sqrt{a_B d}$ for the screening radius $r_s$ of non-relativistic 2D electron gas in a TI thin film of thickness $d$.
This result can be derived by substituting the non-relativistic DOS $g= m/\hbar^2$ into $r_s = \sqrt{\kappa d/e^2 g}$~\cite{huang2021a}.
(Note that the expression of $r_s$ looks like the one in 3D bulk semiconductor, because of peculiar electrostatics of TI film with large dielectric constant).
Eq.~\eqref{eq:comp_eq_ti} gives the same expression of $n_c$ as Eq.~\eqref{eq:n_c}, so the left boundary of the TI phase diagram is exactly the same as in Fig.~\ref{fig:phase1}. The only difference between TI diagram and Fig.~\ref{fig:phase1} is that the distance between $N_c$ and $N_0$ is smaller on the right hand side of the TI phase diagram. 
%This is why we are not showing $n$-$N$ plane phase diagram of the TI film here.

\section{Narrow gap completely compensated three-dimensional semiconductor}
\label{sec:3D}

In this section we return to the case of a three-dimensional (3D) completely compensated semiconductor with a gapped Dirac dispersion, from which we started the Introduction section. 
This problem was studied previously~\cite{shklovskii1972,shklovskii1984,skinner2012} for relatively large gap semiconductors, when the random potential is screened by small and well-separated electron and hole droplets, as depicted in Fig.~\ref{fig:large_gap} (see also Fig.~13.4 of  Ref.~\cite{shklovskii1984}). But the recent interest in three-dimensional Dirac semimetals has brought renewed emphasis on 3D materials with a Dirac-like dispersion and very small gap. Such a dispersion arises ubiquitously near the boundary between topological and trivial insulator phases. Examples include ZrTe$_5$ and HfTe$_5$ \cite{Weng2014}, BiTeI \cite{Tran2014}, BiTeBr \cite{Ohmura2017}, and BiSb \cite{Singh2016, Vu2021}. In these materials the gap $\Delta$ can be controlled by external parameters like pressure or magnetic field. 

Below we address both cases $\Gamma_1 \ll \Delta$ and $\Gamma_1 \gg \Delta$. Here our notation $\Gamma_1$ replaces $\Gamma$ of previous sections because in the 3D strong disorder case the self-consistent amplitude of the potential is somewhat smaller than for 2D. According to the self-consistent theory of Ref.~\onlinecite{skinner2014}, the amplitude of the disorder potential at $\Delta = 0$ follows
\begin{align}\label{eq:gamma1}
    \Gamma_1 = \frac{e^2 N^{1/3}}{\kappa \alpha^{1/2}}.
\end{align}
As in Sec.~\ref{sec:lowdisorder}, below we use the notation $\Gamma_1$ only to represent the impurity concentration $N$. In the limit $\Gamma_1 \ll \Delta$ the actual amplitude of the random potential is of order $\Delta$. We also define the characteristic concentration $N_1 = \alpha^{3/2}e^{-6}\kappa^3\Delta^3$, such that $\Gamma_1/\Delta = (N/N_1)^{1/3}$. 

Below we briefly discuss the temperature dependence of the conductivity at $\Gamma_1 \ll \Delta$, which is similar to the 2D case with $\Gamma \ll \Delta$, discussed in Sec.~\ref{sec:lowdisorder}. We then consider the conductivity in the case when the amplitude of the random potential $\Gamma_1 \gg \Delta$, and we emphasize the dramatic difference between 2D and 3D in this case. 

In the case $\Gamma_1 \ll \Delta$, a 3D completely compensated semiconductor is an insulator with three low temperature mechanisms of conductivity. At relatively high temperature electrons and holes can be activated from the Fermi level to their percolation levels. Because in 3D percolation requires only a small fraction of space, $\sim 17\%$, this activation energy is relatively small~\cite{skinner2012}, $E_a \approx 0.3 \Delta$.

At lower temperatures activation to the percolation level is replaced by NNH, with activation energy given by the charging energy of a droplet, $E_C = e^{2}/\kappa R$, where the droplet radius $R= a_B (Na_B^3)^{-1/9}$ for $Na_B^3\gg 1$ or $N/N_1 \gg \alpha^{9/2}$ and $R=a_B$ for $Na_B^3 \ll 1$ or $N/N_1 \ll \alpha^{9/2}$. This leads to $E_c=\alpha^{3/2} \Delta (N/N_1)^{1/9}$ for $N/N_1 \gg \alpha^{9/2}$ and $E_c = \alpha^2\Delta$ for $N/N_1 \ll \alpha^{9/2}$. 

The prefactor of NNH conductivity is equal to $\exp(-S)$, where using Eq.~\eqref{eq:big_lambda} we get 
\begin{align} \label{eq:act}
    S= \frac{\Lambda \Delta}{\hbar v} = \frac{1}{\alpha^{1/2}} \left(\frac{\Delta}{\Gamma_1}\right)^3 = \frac{N_1}{\alpha^{1/2} N}.
\end{align} 

At the lowest temperatures the conductivity is dominated by ES VRH. 
Using Eqs.~\eqref{eq:big_lambda}, ~\eqref{eq:xi_d}, and ~\eqref{eq:act} we get $ T_{\rm ES} = e^2/\kappa \xi = \alpha \Delta$.

Let us now turn to the conductivity in the case $\Gamma_1 \gg \Delta$, for which the bending of conduction and valence bands looks similar to Fig.~\ref{fig:puddles}.
In this case electrons occupy almost half of space, but we know that only 17\% of the space is enough to provide percolation. 
This means that in the 3D case, in contrast with the 2D case discussed in Section \ref{sec:fractal}, at $\Gamma_1 \gg \Delta$ we deal with a good metal.
There is therefore a critical disorder strength, such that $(\Gamma_1/\Delta)_c \sim 1$, which produces a percolative insulator-to-metal transition (IMT).

Assuming that the random potential energy $u(\vb{r})$ follows the Gaussian distribution, we can estimate $(\Gamma_1/\Delta)_c$ and the critical concentration $N_c$ of IMT by equating the volume fraction of electron (or hole) puddles to 17\%.
Since the condition to have hole puddles is $u > \Delta$ (c.f. Fig.~\ref{fig:large_gap}), we have
\begin{align}
    \int_{\Delta}^{\infty} \frac{e^{-u^2/2\Gamma_1^2} du}{\sqrt{2\pi}\Gamma_1}=0.17. 
\end{align}
By solving this equation we get the critical $(\Gamma_1/\Delta)_c = 1.05$ and the corresponding impurity concentration $N_c = 1.15 N_1$ at the IMT.

One can estimate the small width of this IMT by calculating the hopping conductivity at the vicinity of IMT where $(\Gamma_1/\Delta)_c - \Gamma_1/\Delta\ll 1$ and finding the distance from the IMT where $[(\Gamma_1/\Delta)_c - \Gamma_1/\Delta]$ ceases to be exponentially small. This estimate would remind the calculation of the width of integer QHE steps~\cite{polyakov1993}. Such a theory is beyond the scope of this paper.
\begin{figure}[t]
    \centering
    \includegraphics[width=\linewidth]{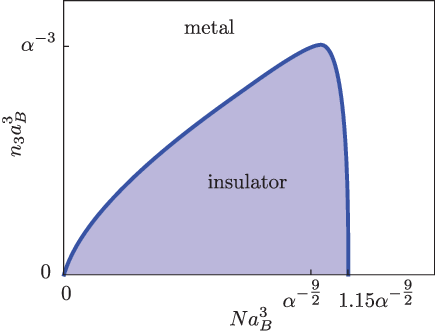}
    \caption{Schematic $n_3$-$N$ plane phase diagram of a 3D narrow gap strongly compensated semiconductor. The shaded blue domain is the insulator phase while the white domain is the metal phase. The phase boundary follows Eq.~\eqref{eq:n_3c} on the left side, reaches the maximum $n_3 a_B^3=\alpha^{-3}$ near $N a_B^3 = N_1 a_B^3 = \alpha^{-9/2}$, and then vertically drops at $N = N_c =1.15 N_1$.} 
    \label{fig:phase2}
\end{figure}

So far in this section we dealt with completely compensated semiconductor. Let us now briefly consider strongly compensated $n$-type semiconductor in which concentrations of donors $N_D$ and acceptors $N_A$ are close, but different. It is convenient to describe such a semiconductor by the total concentration of Coulomb impurities  $N = N_D + N_A$ and three dimensional concentration of electrons $n_3 = N_D - N_A \ll N$. We are interested in the phase diagram of IMT in the $n_3$-$N$ plane, which is
shown in Fig.~\ref{fig:phase2}.

At modest impurity concentration $1< N a_B^3 < N_1 a_B^3 = \alpha^{-9/2}$, we equate the Fermi energy with the Coulomb potential energy fluctuations to find the critical concentration $n_{3c}$
\begin{align}\label{eq:comp_eq_3}
    \frac{\hbar^2 n_3^{2/3}}{m} =  \frac{e^2\sqrt{N r_s^3}}{\kappa r_s},
\end{align}
where $r_s = a_B^{1/2} n_3^{-1/6}$ is the screening length for non-relativistic 3D electron gas.
By solving Eq.~\eqref{eq:comp_eq_3} one obtains~\cite{shklovskii1984,shklovskii1971a}
\begin{align}\label{eq:n_3c}
    n_3=n_{3c} = N^{2/3}/a_B,
\end{align}
which is shown by the small  $N$ side phase boundary in Fig.~\ref{fig:phase2}.
Near $N=N_1$ the phase boundary drops almost vertically to $n_3=0$ at $N=N_c = 1.15 N_1$. Thus, as in 2D case we arrive at disorder driven reentrant metal-insulator-metal transition. Note that for 3D case, we use the word ``metal'' as opposite to 2D term ``almost metal'' we used in Fig.~\ref{fig:phase1}.
This is because real metallic phase is allowed only in 3D.

\section{Summary and Conclusion}
\label{sec:conclusion}

In this paper we have considered the temperature-dependent conductivity of a two-dimensional insulator subjected to disorder by Coulomb impurities in the substrate.
%, with equal concentrations $N/2$ of both sign.
Our primary results can be summarized as follows. When the impurity concentration $N$ is below a certain value $N_0$ [see Eq.~(\ref{eq:N0})], the random potential of charged impurities necessarily produces large band bending, which the amplitude
$\Gamma$ becomes much larger than $\Delta$. Then the system can be described as a network of large and closely-spaced fractal puddles  [Fig.~\ref{fig:puddles}] separated by narrow insulating barriers [Fig.~\ref{fig:fingers}]. This disorder landscape implies low-energy pathways for electron conduction, leads to the ``three-mechanism sequence'' illustrated in Fig.~\ref{fig:conductivity}. The high temperature regime with $E_a = \Delta$ is relegated to only such high temperatures that $T$ is comparable to $\Delta$. The second regime, the nearest neighbor hopping between puddles (NNH), exhibits a parametrically smaller activation energy, whose value depends on the impurity concentration. At the lowest temperatures the conductivity is due to the Efros-Shklovskii variable range hopping (VRH), which may appear as an even smaller activation energy when measured over a limited temperature range. Experiments are instead more likely to observe NNH or ES VRH, with an activation energy that declines very rapidly with increasing $N$ [Fig.~\ref{fig:energy_n}]. 

When the impurity concentration $N$ exceeds another critical value $N_c$ 
%(which is remarkably close to  $N_0$ [Fig.~\ref{fig:energy_n}]),
the tunnel barriers between puddles become thin enough to be nearly transparent, and electrons are delocalized across many puddles. In this limit the conductivity follows ES law with the localization length growing exponentially with increased disorder. The corresponding apparent activation energy falls exponentially, so that in mesoscopic samples one effectively has an unconventional disorder-induced insulator-to-metal transition. The analogous problem for three-dimensional insulators (see Sec.~\ref{sec:3D}) shows a genuine IMT due to percolation of electron and hole puddles separately. 

Above we were talking about the neutrality point. When the Fermi level is away from neutrality point and the concentration of impurities is relatively small, there is a conventional metal-insulator transition with increasing disorder. Combining it with insulator-metal transition at large impurity concentrations away from neutrality we arrive at a disorder driven re-entrant metal-insulator-metal transition. (See phase diagrams of such transitions shown in Fig.~\ref{fig:phase1} and Fig.~\ref{fig:phase2}.)

Our results have implications for a wide variety of experiments on 2D electron systems with a narrow energy gap. Some of these include 2D and thin 3D TIs, Bernal bilayer graphene with a perpendicular displacement field, and twisted bilayer graphene, as mentioned in the Introduction. In such systems the temperature-dependent conductivity is often used as a primary way to diagnose the magnitude of energy gaps. Our results here suggest that such studies suffer an essentially unavoidable limitation, since the apparent activation energy $E_a$ at low temperature has no simple relation to the energy gap, and in general $E_a$ can be taken only as a weak lower bound. No wonder that the transport activation energy in many cases is 10-100 times smaller than the value expected theoretically or measured by probes like optical absorption or tunneling spectroscopy. In this paper we studied in details gapped thin films of 3D topological insulators, which due to the large dielectric constant have peculiar 3D-like electrostatics (see Sec.~\ref{sec:TI}).

The existence of an apparent disorder-induced IMT in strongly compensated semiconductor is an especially striking result of our analysis. For conventional insulators, this apparent transition cannot be called a true IMT, since in 2D the zero-temperature conductance flows toward zero in the thermodynamic limit for any finite amount of disorder \cite{Abrahams1979}. However, the situation may be different for thin TI films, since the spin-orbit coupling of the TI surface states permits a stable metallic phase \cite{Hikami1980, Mong2012}. A full theory of this IMT in TI films is beyond the scope of our current analysis.

Above we discussed how conductivity changes with the growth of the concentration of impurities $N$. 
However, in a typical experiment $N$ is not well known and remains fixed, while $\Delta$ is tuned. 
For example, in BLG this tuning is done by changing orthogonal displacement field $D$. 
Then our theory implies that measured at high temperature activation energy $E_a$ agrees with the value of the gap 
$\Delta(D) \propto D$ predicted by Ref.~\onlinecite{mccann2013,Slizovskiy2021}, if roughly $\Delta(D)  > \Gamma = \alpha^{-2/3} e^2 \kappa^{-1} N^{1/3}$.
On the other hand, if $\Delta(D) < \Gamma$ then the observed activation energy $E_a$ is much smaller than the gap $\Delta(D)$.
Thus, measuring minimum value of $E_a$ coinciding with theoretical $\Delta(D)$ one can find $N$. 
Let us do this for the studied in Ref.~\onlinecite{icking:2022} sample of BLG separated by thin hexagonal boron nitride (hBN) layers from silicon oxide, assuming that the majority of charged impurities are located in the bulk of SiO$_2$. The smallest observed $E_a(D)=11$ meV still resides on the theoretical line $\Delta(D)$ [c.f. Fig. 3 (c) in Ref.~\onlinecite{icking:2022}]. 
This means that $N \leq N_0(E_a)$. Using the Dirac velocity for graphene $v \sim 10^6$ m/s for bilayer graphene layers with SiO$_2$ $\kappa = 4$, we get $\alpha \simeq 0.5$. For $E_a(D)=11$ meV we get $N \leq 10^{16}$cm$^{-3}$.

Let us now estimate the concentration of charged impurities for BLG samples sandwiched between two hBN layers and deposited on the graphite gate~\cite{icking:2022} (they are likely C atoms~\cite{onodera:2020} substituting for B and N). For this sample $E_a$ agrees with theoretical $\Delta(D)$ till $E_a = 0.5$ meV. This means that in hBN $N$ is so small that impurities are distributed in the layer of thickness $d$ much smaller that $N^{-1/3}$. We generalized our theory to this case and arrived at the estimate $N \leq  2 \times 10^{14}$ cm$^{-3}$ for the state-of-the-art hBN. Indeed, for impurities located in a thin hBN layer of thickness $d \ll N^{-1/3} < \lambda$ the potential mean-square fluctuation 
\begin{align*}
    \Gamma^2 \simeq (e^4 N d/\kappa^2) \ln(\lambda\sqrt{Nd}).
\end{align*}
Combining with Eq.~\eqref{eq:lambda1}, we get 
\begin{align*}
    \lambda = \alpha^{-12/7} (Nd)^{-1/2}, \,\, \Gamma = \alpha^{-2/7} e^2 \sqrt{Nd}/\kappa.
\end{align*}
Using $E_a \simeq 0.5$ meV and $d \simeq 20$ nm, we get that in hBN $N\leq N_0 (E_a) = \alpha^{4/7} \kappa^2 E_a^2/e^4 d \simeq 2\times 10^{14}$ cm$^{-3}$. 

%Finally, we mention that the tunneling between puddles can be reduced by applying a magnetic field orthogonal to the 2D plane. This reduction leads to an exponential positive magnetoresistance in the insulating phase, similar to the one studied in  Ref.~\cite{shklovskii1984}. The theory of such magnetoresistance is also beyond the scope of this paper.

\begin{acknowledgments}
We are grateful to David Goldhaber-Gordon, Ilya Gruzberg, Shahal Ilani, Kin Fai Mak, Koji Muraki, Stevan Nadj-Perge and Christoph Stampfer for helpful discussions.
Y.H. gratefully acknowledges support from Larkin Fellowship at the University of Minnesota. 
B.S. was partly supported by NSF grant DMR-2045742.
\end{acknowledgments}

\end{document}